\documentclass[%
 aip,
 amsmath,amssymb,
preprint,%
]{revtex4-2}
\usepackage{hyperref}
\usepackage{graphicx}
\usepackage{dcolumn}
\usepackage{bm}
\usepackage{ulem}
\usepackage{comment}
\usepackage{cancel}
\usepackage{xcolor}
\usepackage{mathrsfs}
\usepackage{amsfonts}
\usepackage{amsmath}

\begin{document}


%
%

\title{Compressible helical turbulence: Fastened-structure geometry and statistics}
\author{Jian-Zhou Zhu
}
 \email{jz@sccfis.org}
\affiliation{$^1$Su-Cheng Centre for Fundamental and Interdisciplinary Sciences, Gaochun, Nanjing 211316, China and Fluid Institute, Nanchang University, Nanchang 330031, China.
}

\date{???
}

\begin{abstract}
Reduction of flow compressibility with the corresponding ideally invariant helicities, universally for various fluid models of neutral and ionized gases, can be argued statistically and associated with the geometrical scenario in the Taylor-Proudman theorem and its analogues. A `chiral base flow/field', rooted in the generic intrinsic local structure, as well as an `equivalence principle' is explained and used to bridge the single-structure mechanics and the helical statistics. The electric field fluctuations may similarly be depressed by the (self-)helicities of the two-fluid plasma model, with the geometry lying in the relation between the electric and density fields in a Maxwell equation.
\end{abstract}

\maketitle

\section{Introduction}\label{sec:introduction}
Compressible fluid models are the appropriate choice and useful in describing the dynamics of neutral and ionized gases when the kinetic effects are negligible and when incompressibility approximation is inaccurate. Lots of nice mathematical structures have been exposed and powerful topological and geometrical tools developed for hydrodynamics (HD).\cite{ArnoldKhesin98Book,MorrisonRMP1998} However, the study of what a (controllable) quantity or property of the flow itself influences or controls the (degree) of the compressibility of the flow, and then the physical consequences, appears still much wanted. For example, (in)stability, relaxation and heating are important topics of laboratory (especially fusion research) and space-/astro-physical plasma studies on issues related to compression of flows and/or with compressible fluid models,\cite{MorseNFtext2018,
LorenziniNP2009,SrinivasanShumlakPoP2011} and, among many other relevant works (e.g., Refs. \onlinecite{MatthaeusGoldsteinJGR1982StriblingMatthaeusGhoshJGR1994,
BalsaraPouquetPoP1999,ChristenssonHindmarshPRE2001,SagautCambon2008Book,Schlickeiser2002Book,SteinhauerIshidaPoP1998}), it has been particularly concluded \cite{Schlickeiser2002Book} that cross and magnetic helicities can influence the acceleration rate of charged particles, but the roles of compressibility and helicity have not been clear enough, due to the possible limitations of the analytical (Fokker-Plank phenomenology \cite{DungSchlickeiserAA1990ab}) and numerical model (incompressible magnetohydrodynamics --- MHD \cite{TeacaEApre2014}): Compression is associated to the first-order Fermi process and the electric-field to the second-order one, then we can naturally ask whether and how dynamically (from the first principle) the helicity affects the compressibility and then the heating, as partially addressed in Ref. \onlinecite{Z16} where the internal energy (the work part of the pressure) is equivalently described by the density fluctuations with the adiabatic relation used (see below).

Another specific example is that the prediction of inverse magnetic helicity cascade was made from the incompressible MHD absolute equilibrium analysis,\cite{FrischJFM75} key to modern nonlinear dynamo theory, but the robustness with respect to the compressibility effect has not been considered. It is not impossible that compressibility may change the picture. For instance, it can be argued \cite{Z16} that the three-dimensional inverse energy cascade numerically demonstrated\cite{bmt12} (see also Ref. \onlinecite{WaleffePoF92}) for incompressible homochiral turbulence with positive definite helicity should not present in the compressible setting. Indeed, while most studies of helicities relevant to plasma turbulence are on transport and dynamo issues, much less attention has been paid to the direct consequence on the \textit{flow compressibility} property, to our best knowledge. Some numerical simulations \cite{BalsaraPouquetPoP1999} were performed with the compressible model, thus in a sense confirmed the robustness, but with no consideration or clarification of the compressibility effect; similarly, the stability analysis\cite{WaleffePoF92,LinkmannJFM2015} has not been well demonstrated in the compressible situation, with the longitudinal and transverse modes being dynamically interacting. It will be more satisfying to theoretically address the problems with some unity in the treatment, with consistent conservation laws\cite{BanerjeeGaltierPRE2013}. Note that the notion of `flow compressibility' in this note is not that defined by the variation of density with respect to pressure, $\partial \rho/\partial p$, through the state equation, the latter of a gas being not indicated to be directly changed when we say ``compressibility reduction'': We mean, even if the state equation is fixed, the strength of the fluctuations of velocity divergence and of density is (relatively) reduced.

It is indicated \cite{Z16} with the calculation extending that of \citet{K55} by respecting the helicity that the latter may reduce the HD compressibility (`fasten the flow'), and the preliminary numerical analysis \cite{HTG} appears to be consistent with the indication even reasonably beyond the postulated conditions for analytical calculations in Ref. \onlinecite{Z16}. Actually it is possible to show that some recent numerical result of the Gross-Pitaevskii equation modeling the low-temperature superflow or Bose-Einstein condensate \cite{ClarkDiLeoniMininniBrachetPRA16} is also consistent with the `fastening' notion. Such statistical mechanics analyses applied to various fluid models \cite{K73,FrischJFM75,Z16,MiloshevichLingamMorrisonNJP17} however have not been mechanically associated with the geometry and/or topology of the turbulence structures;\cite{Moffatt78Book,ArnoldKhesin98Book} for recent advancements, particularly for the geometry and topology of extended magnetohydrodynamics (MHD), c.f., Refs. \onlinecite{LingamMiloshevichMorrisonPLA16,BesseFrischJFM17}; and, for two-fluid plasma model (TFPM), see, e.g., Refs. \onlinecite{TurYanovskyJFM93,SagdeevMoiseevTurYanovskii86,ZhuPoF18Lie}), which is the bigger point we also try to make, towards a theory of statistical geometrical fluid mechanics and a more fundamental approach to the relevant applied physical issues, such as the cosmic ray acceleration and flow control (ranging from, say, civil engineering to fusion). 

Related to the united effects of both compressibility and (magnetic/cross-)helicity in magnetohydrodynamics, there have been many simulations, with, for example, the observation of the `breakdown' of cross-helicity together with the increase of compressibility (conjectured to be the spontaneous generation of magnetosonic `structures') \cite{ServidioPaSS2007} when the Hall term dominates the simulated slice of magnetopause boundary layer in the equatorial plane, among other different interests,\cite{BalsaraPouquetPoP1999,ChoLazarianTCFD2005,YangLiMNRAS2019} but deriving the  information about the partitions between the longitudinal and transverse (with respect to the wavevectors) modes requires extra efforts. The energy transfer rates in compressible two-fluid plasmas have been studied,\cite{BanerjeeAndresPRE2020} but helicities were considered only in the incompressible limit.\cite{AndresGaltierSahraouiPRE2016} Thus, we will also go beyond XMHD to such compressible TFPM \cite{SteinhauerIshidaPoP1998} for a more comprehensive understanding of the helicity effects. Note that recently $^3$He-rich solar energetic particle events associated to type III radio bursts and helical jets have been observed at the Parker Solar Probe and near Earth,\cite{WiedenbeckEAapjs2020} and a theory explicitly with electric fields and flow structures would be attractive.

\section{Statistical geometrical fluid mechanics
}\label{sec:CBFrot}
To prepare for later discussions, we first briefly review the early approach to aeroacoustics of Kraichnan \cite{K55} which is recently updated for compressible helical turbulence.\cite{Z16} An ensemble theory is then proposed with the elements of a \textit{chiral base field/flow}, an \textit{equivalence principle} and a \textit{fastening notion} for the compressibility reduction by the corresponding helicities, in accordance with the indications from the absolute equilibrium results, universally for various fluid models of neutral and ionized gases.

\subsection{\label{sec:statisticsRetrospect}Basic and fundamental statistical analysis in retrospect}
Kraichnan\cite{K55} let $\bm{u}$, $\rho$, $\eta$ and $c$ represent the velocity, density, viscosity and velocity of sound in a compressible flow, with the pressure $p$ given by the adiabatic relation 
\begin{eqnarray}
\text{$p=c^2\rho$ and $\rho=\rho_0 e^\zeta$ with an equilibrium $\rho_0$}.\label{eq:adiabatic}
\end{eqnarray}
In this note, both $c$ and $\rho_0$ are unit ($=1$), with appropriate choice of scales and units, but sometimes we still let them present explicitly to be more physically sensible. The equations of motion then write
\begin{eqnarray}
  \partial_t \zeta+\zeta_{,\alpha}u_{\alpha}+u_{\alpha,\alpha} &=& 0, \label{eq:k1}\\
  \partial_t u_{\lambda} + u_{\sigma}u_{\lambda,\sigma}+c^2\zeta_{,\lambda}-\eta \theta_{\lambda\sigma,\sigma}&=& 0, \label{eq:k2}
\end{eqnarray}
where $\theta_{\alpha \beta}=u_{\alpha,\beta}+u_{\beta,\alpha}-\frac{2}{3}\delta^{\alpha}_{\beta}u_{\sigma,\sigma}$, $(\bullet)_{,\gamma}=\partial (\bullet)/\partial x^{\gamma}$. 
Working in a cyclic box of dimension $2\pi$ with $V=[0,2\pi)^3$ and applying Fourier representation for all the dynamical variables $v(\bm{r}) \to \hat{v}(\bm{k})$, say, $\bm{u}(\bm{r})=\sum_{\bm{k}}\hat{\bm{u}}(\bm{k}) \exp\{\hat{i}\bm{k}\cdot \bm{r} \}$ with $\hat{i}^2=-1$, he constructs a phase space by the real and imaginary parts of $v$s and showed that the flow in such a phase space is incompressible, the Liouville theorem, in the inviscid case.
Kraichnan takes the `small-excitation' approximation for the ideal invariant of total, kinetic plus
\begin{equation}\label{eq:zeta}
\!\!\!\! \text{internal/potential ($\int_{\rho_0}^{\rho} \frac{p-p_0}{\rho^2} d\rho 
 \approx \frac{c^2 \zeta^2}{2}$ for small $\zeta$)}
\end{equation}
mean energy per unit mass: 
\begin{equation}\label{eq:E}
\mathcal{E}=\frac{\langle u^2 + c^2\zeta^2 \rangle}{2} = \frac{ \sum_{\bm{k}}[\hat{u}_{\lambda}(\bm{k})\hat{u}^*_{\lambda}(\bm{k})+c^2|\hat{\zeta}(\bm{k})|^2] }{2}
\end{equation}
with $\langle \bullet \rangle=\frac{1}{(2\pi)^3}\int_V \bullet dV$. $\mathcal{E}$ is not rigorously an ideal (inviscid) invariant; but Kraichnan used it to approximate the invariant total energy. Galerkin truncation, say, imposing all modes with $k=|\bm{k}|$ greater than some cut-off value $K$ to be zero, is then performed, which does not change the approximated conservation of energy due to its being quadratic and diagonal in $\bm{k}$.\cite{K73} Reasoning with the $H$-theorem, Kraichnan expects an ensemble of systems tend towards an absolute statistical equilibrium state. We remark that even if the Galerkin truncated absolute equilibria were not fully realizable for various reasons, they may still be valuable because the other actors (forcing and damping etc.) in turbulence may actually help the (quadratic) interactions to mix up the phase space and thus the  hypothesized (partial) thermalization to the absolute equilibria being still relevant in appropriate turbulence situations. 

However, as in the incompressible case \cite{K73}, the helicity 
\begin{equation}\label{eq:H}
\text{$\mathcal{H} := 
\frac{\langle \bm{\omega} \cdot \bm{u} \rangle}{2}$, with the vorticity $\bm{\omega}=\nabla\times\bm{u}$},
\end{equation}
is also a working ideal barotropic invariant.\cite{Moffatt78Book} 

The Helmholtz decomposition theorem can be `sharpened' by further decomposing the transverse (`solenoidal') velocity field into a left-handed and right-handed chiral modes of sign-definite helicity intensity \cite{WaleffePoF92}
\begin{equation}\label{eq:hd}
\hat{\bm{u}}(\bm{k})=\hat{u}_+(\bm{k})\hat{\bm{h}}_+(\bm{k})+\hat{u}_-(\bm{k})\hat{\bm{h}}_-(\bm{k})+\hat{u}_|(\bm{k})\bm{k}/k,
\end{equation}
with $\hat{i}\bm{k}\times \hat{\bm{h}}_s=sk\hat{\bm{h}}_s$ (for $\hat{i}^2 = -1$) and $s=\pm$, and that, 
\begin{eqnarray}
\mathcal{E}=\frac{1}{2}\sum_{\bm{k}}|\hat{u}_+|^2+|\hat{u}_-|^2+|\hat{u}_||^2+c^2|\hat{\zeta}|^2,\label{eq:ce}\\
\mathcal{H}=\frac{1}{2}\sum_{\bm{k}} k|\hat{u}_+|^2-k|\hat{u}_-|^2.\label{eq:ch}
\end{eqnarray}
Following Kraichnan's \cite{K55} procedure, we obtained with the canonical distribution $\sim \exp\{-(\alpha \mathcal{E} + \beta \mathcal{H})\}$ the  absolute-equilibrium modal spectra \cite{Z16}
\begin{eqnarray}
  U^{\pm}_K(\bm{k}) := \langle |\hat{u}_{\pm}|^2 \rangle &=& \frac{1}{\alpha\pm\beta k}, \label{eq:u+}\label{eq:Z16Upm}\\
  Z(\bm{k}) := \langle c^2|\hat{\zeta}|^2 \rangle &=& \frac{1}{\alpha} =\langle |\hat{u_|}|^2 \rangle =: U^|_K(\bm{k}). \label{eq:VA}
\end{eqnarray}
With vanishing $\beta$ or $\mathcal{H}$, these spectra present equipartition, just as Lee's \cite{Lee1952,hydrochirality} earlier incompressible result and that of Kraichnan.\cite{K55} As remarked in the introductory discussions, it was observed\cite{Z16} that $\alpha$ in Eq. (\ref{eq:VA}) should be positive, thus no concentration at large scales for the homochiral version of Eq. (\ref{eq:Z16Upm}) with negative $\alpha$, and thus no indication of three-dimensional inverse energy cascade as that for the incompressible model.\cite{WaleffePoF92,bmt12,hydrochirality}

For the application of the absolute equilibrium analysis besides the most conspicuous one of reasoning the cascade/transfer direction such as that of magnetic helicity,\cite{FrischJFM75} it deserves to digress to highlight the first  usage by Lee\cite{Lee1952} himself who remarks: 
\begin{quote}
In the case of a real fluid ... be represented by a constant energy flow from small wave number regions to large wave number regions ... this energy transfer makes it necessary that the energy spectrum must be quite different from the equilibrium case ... since this energy transfer is only between different $|\bm{k}|$ values, it is expected that all modes of Fourier components of the same $|\bm{k}|$ values may be still in energy equipartition.
\end{quote}
Thus, he has explained the mechanism for the \textit{isotropy} notion which was only `hypothesized' by Kolmogoroff. He has also predicted energy equipartition between the magnetic and kinetic fields for MHD turbulence with neither magnetic nor cross helicity (see below), and states
\begin{quote}
Similar arguments also lead to the conclusion that both magnetic field and velocity field are isotropic in the Kolmogoroff region.
\end{quote}
These conclusions appear inconspicuous, because they were long hypothesized and used a lot, but obviously deep and hefty. Similarly, Kraichnan\cite{K55} has not made a very pronounced discovery for compressible turbulence, but his analysis of the connection with real turbulence is also fundamentally innovative. We know that both Lee's and Kraichnan's spectra can be enriched by including the constraints of corresponding helicities\cite{K73,FrischJFM75,Z16} and can be refined by applications of helical decomposition,\cite{hydrochirality} which means that the absolute equilibrium analysis is still calling for keen applications and creative developments, thus old-fashioned but forever young.   

Now, the solenoidal-mode energy becomes larger than the `acoustic' one in the helical case,
\begin{equation}\label{eq:neq}
U^{\perp}_K := U^{+}_K + U^{-}_K > Z+U^|_K=: U^{\sim}_K.
\end{equation}
Comparing the helical and nonhelical cases, one sees that the interactions for thermalization with helicity tend to reduce the energy proportions of $Z$, $U^|_K$ and $U^{\sim}_K$, i.e., lowering the energy of `noise' (in the aeroacoustic context), which is the basic observation used for physical inference in Ref. \onlinecite{Z16}, a kind of \textit{comparative} analysis different to previous ones\cite{K73,FrischJFM75,hydrochirality} mainly based on the helical absolute equilibria themselves. Such statistical dynamical (thermalization) mechanism and consequent property may be expected to persist for turbulent flows in which the energy damping and injection mechanisms are independent of helicity, but the discussions for cases with forcing and dissipation, where other effects enter (to assist or resist the mechanism) and other phenomena mingle (to increase or decrease the property), should of course be careful, as performed in Ref. \onlinecite{Z16}. That is, Eqs. (\ref{eq:u+},\ref{eq:VA},\ref{eq:neq}) and their persistent residual in turbulence mean the reduction by helicity of the compressive, density and potential/internal-energy modes, thus of the `compressibility' of flows. Or, in other words, the compressible helical turbulence may be `fastened', which is to be further understood mechanically/geometrically below, in a sense of statistical geometrical fluid mechanics. 

Other thermalization residues, such as that related to the ``strong acoustic equilibrium'', \textit{i.e.}, the equipartition between the dilatational/compressive mode and density mode --- also the pressure-potential/internal-energy mode for the particular model analyzed here --- as presented in the above spectra, helical or not, can also be accordingly discussed;\cite{Z16} and, still more are to be discovered with appropriate combination of tradition and innovation.\cite{traditon&innovation} 

While the approach outlined in the above indeed helps us thinking coherently, instead of being in a panic, when confronted with a turbulence, the prediction, such as that from comparing Eq. (\ref{eq:neq}) with Kraichnan's\cite{K55} nonhelical equipartition one, for real turbulence is, strictly speaking, a semi-heuristic; thus, further or complementary investigations must be performed as follows.

\subsection{\label{sec:geometry}
\textit{Chiral base flow/field}, mean rotation and guiding fields, and, the equivalence principle of helical turbulence}

Central to the rotating HD is the Taylor-Proudman effect (TPE) which, for an incompressible flow (\textit{e.g.}, \citet{Chandrasekhar61Book}), states the asymptotic incompressible and two-dimensional (2D) `horizontal' velocity field $\bm{u}_h$ in the rotating plane, thus also 2D `vertical' component $\bm{u}_v$ with the incompressible $\bm{u}_h + \bm{u}_v$. If the flow is compressible, $\bm{u}_v$ is three-dimensional (3D), which can be shown by extending straightforwardly the standard `dominant-balance-fashion' proofs, analytically or geometrically\cite{Zhu19TaylorProudman} (see also Appendix \ref{app:MHD} for the relevant remark of the compressible MHD analogue). [The geometrical proof will be applied below to the chiral base flow, to be introduced, to demonstrate the picture of our `fastening' notion.] Thus, the compressible Taylor-Proudman limit results in a two-component-two-dimensional coupled with one-component-three-dimensional (2C2Dcw1C3D) $\bm{u}$ whose spatial gradient matrix is \textit{uniformly} of the \textit{real Schur form} or \textit{real Schur flow} (RSF: \textit{c.f.} Ref. \onlinecite{Zhu18PoF2C2Dcw1C3D} and references therein, as well as Ref. \onlinecite{LiZhangHeAPS2014} for applications of \textit{local} canonical RSF in kinetic analyses of flow patterns). The component $\bm{u}_h$ however can be dilatational in the horizontal plane for a general RSF which can also have nonvanishing helicity (`helical').

\begin{figure}
\begin{center}
\includegraphics[width=15.0cm]{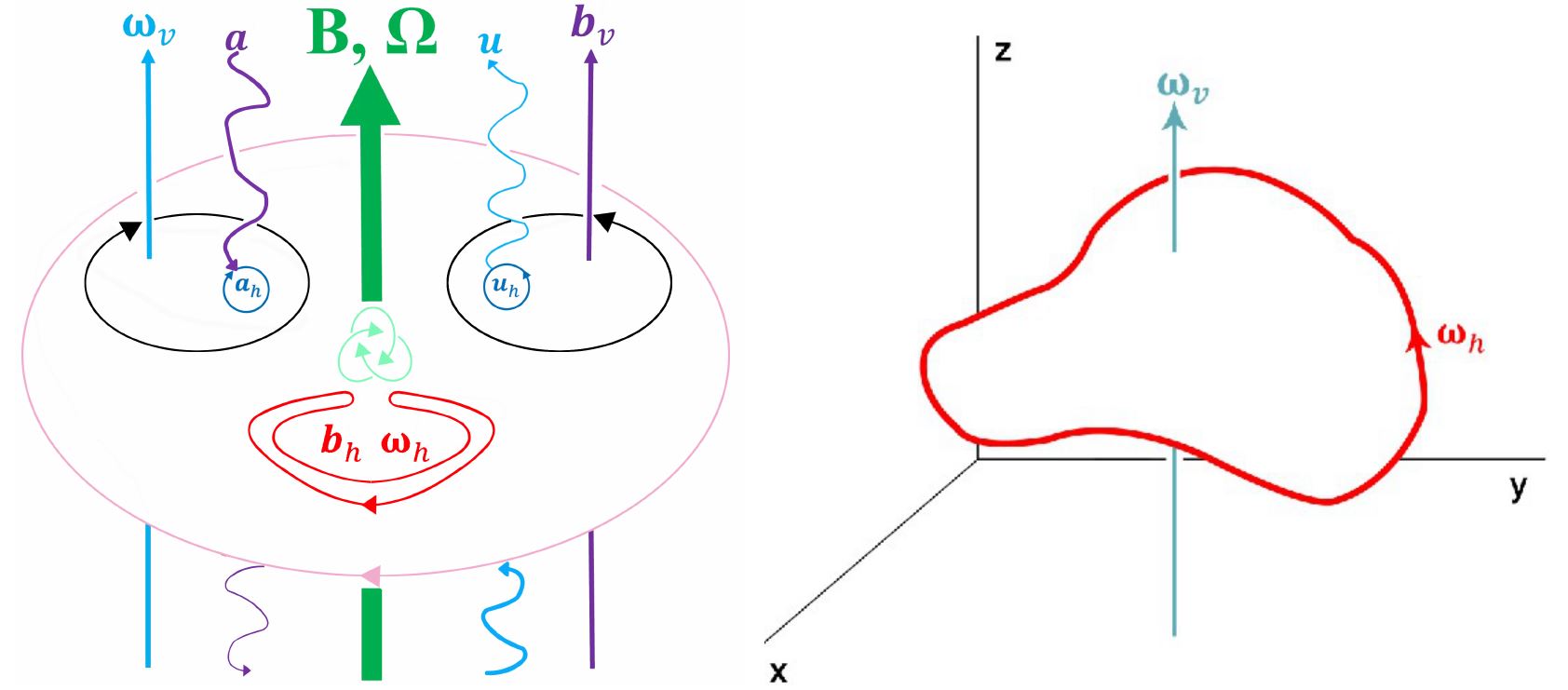}
\end{center}
\caption{The geometry of \textit{chiral base flow/field}: 
Closed lines are the (projected) horizontal vorticity or velocity streamlines, perpendicular to which are the straight vertical ones. Screws denoted by $\bm{u}$ (and $\bm{a}$) are the helical streamlines of velocity (and the vector potential of magnetic field $\bm{b}=\nabla \times \bm{a}$) on some vertical cylindrical surfaces. The velocity and vorticity streamlines, knotted (as symbolized by a trefoil knot) or not, may be closed at the boundary (periodic or at infinity) of or inside the domain. 
Sketched are only some simple configurations which typically represent the fastening scenario with the `ropes' and `screws' of the streamlines. The black and purple circles, enriching the fastening scenario, can be (the projections of) either velocity or vorticity lines, thus no notation explicitly assigned, and extra fields of MHD, as already partly described, are also presented for the case with aligned $\bm{\Omega}$ and $\bm{B}$ (Appendix \ref{app:MHD}). The right panel adapts the schematic diagram for the vorticity structure from Ref. \onlinecite{Zhu18PoF2C2Dcw1C3D}.
}
\label{fig:ou}
\end{figure}

Fig. \ref{fig:ou} is for a helical RSF, termed the \textit{(giant/global) chiral base flow/field} [(g)CBF]: as we will see, a helical RSF may be used for the (infinitesimal) neighbourhood of a point or for some finite spatial region, or even the whole domain, thus needing or not the prefix `(g)' before `CBF' depends on the situation; and, sometimes `(g)' can be neglected without confusion when the situation is self-evident. Such CBF has \textit{uniformly}, in the sub-region or whole domain, $\partial_z \bm{u}_h = \bm{0}$, i.e., $\partial_z u_{x} = \partial_z u_{y} = 0$ with the $z$-axis chosen to be the vertical one, thus the decomposition of the vorticity $\bm{\omega} = \nabla \times \bm{u}$ presents the horizontal component
\begin{eqnarray}
\text{$\bm{\omega}_h = (\partial_y u_{z}, -\partial_x u_{z}, 0) = \nabla \times \bm{u}_v$} \label{eq:omegah}
\end{eqnarray}
lying in the $x$-$y$ plane (loops with arrows) and the vertical component
\begin{eqnarray}
\text{$\bm{\omega}_v = (0,0,\partial_x u_{y} - \partial_y u_{x}) = \nabla \times \bm{u}_h$} \label{eq:omegav}
\end{eqnarray}
along the $z$ axis (straight lines with arrows). Neither $\nabla \cdot \bm{u}$ nor $\partial_z u_z$ is required to vanish, but in the barotropic case, both $\bm{\omega}_v$ and $\bm{\omega}_h$ are still ideally frozen-in to the flow as a straightforward generalisation to the barotropic case is possible from the incompressible case of Ref. \onlinecite{Zhu18PoF2C2Dcw1C3D} (whose Fig. 1 is thus reproduced here in the right panel) with the invariant average (over the volume $\mathcal{V}$) helicity density, or simply the \textit{helicity}
\begin{equation}
\mathcal{H} 
=\frac{1}{\mathcal{V}}\int \bm{\omega}_v \cdot \bm{u}_v d^3\bm{r},\label{eq:2D3Chelicity}
\end{equation}
holding when there is no boundary contribution from integration by parts. 
Now, for given $\mathcal{H}$, we introduce such `mean rotation' rate $\bm{\Omega}$ along the vertical direction that, for the relative velocity $$\bm{u}' = \bm{u} - \bm{\Omega} \times \bm{r} = (u_x + y\Omega, u_y - x\Omega, u_z)$$ [thus $x' = x + y\Omega t$, $y' = y - x\Omega t$ and $z' = z$],  the helicity of $\bm{\omega}' = \nabla \times \bm{u}'$ vanishes, \textit{i.e.,} $$\mathcal{H}' = \frac{1}{2\mathcal{V}}\int \bm{\omega}'\cdot \bm{u}' d^3\bm{r} = 0,$$ with $\bm{\Omega}$ determined, from 
\begin{equation}
\bm{\omega} \cdot \bm{u} - \bm{\omega}' \cdot \bm{u}' = \bm{\Omega} \cdot [2\bm{u}' + (\bm{r} \times \bm{\omega}')] = \bm{\Omega} \cdot [2\bm{u} + (\bm{r} \times \bm{\omega})] \label{eq:hhprime}
\end{equation}
for general $\bm{u}$ and Eqs. (\ref{eq:omegah},\ref{eq:omegav}) for CBF, by
\begin{equation}
\mathcal{H} = \bm{\Omega} \cdot \frac{\int [2\bm{u}_v + \bm{r} \times (\nabla \times \bm{u}_v)] d^3\bm{r}}{2\mathcal{V}}. \label{eq:Omega}
\end{equation}
Our strategy is to reduce the relevant $\mathcal{H}$ effect on $\bm{u} = \bm{u}' + \bm{\Omega}\times \bm{r}$ to the $\bm{\Omega}$ effect on the relative motion with no helicity via ``boost'' to the rotating frame, for the CBF such chosen, as generically existing in turbulence, that the $\bm{\Omega}$ is well-defined, with the integration in Eq. (\ref{eq:Omega}) non-vanishing, say. The requirement of `slow' relative motion by Taylor,\cite{TaylorPRSA21} or the equivalent statement of `fast' rotation, may be partly relaxed: For example, we can add arbitrarily a gradient of potential $\nabla \varphi'$ to $\bm{u}'$ in Eq. (\ref{eq:circulation}) below. 
The relative motion also forms a CBF, with
\begin{equation}
\text{$\nabla \cdot \bm{u}' = \nabla \cdot \bm{u}$; $\partial_{z'}
 \bm{u}'
= \partial_z \bm{u}$,}\label{eq:primefield}
\end{equation}
thus (\ref{eq:omegah},\ref{eq:omegav}) for $\bm{\omega}'$.

In the CBF, for any material circuit $c(t)$ with horizontal projection area $\mathcal{A}(t) = \oint_{c(t)} xdy - ydx$, the circulation
\begin{equation}
 \oint_{c(t)}\bm{u} \cdot d\bm{r} = \oint_{c(t)}\bm{u}' \cdot d\bm{r} + 2\Omega \mathcal{A} \label{eq:circulation}
\end{equation}
is invariant, if the pressure term does not contribute (as in the barotropic case). 
If $\oint_{c}\bm{u}' \cdot d\bm{r}$ varies comparatively slowly, $\mathcal{A}$ changes little, which is the geometrical argument to prove the TPE by \citet{TaylorPRSA21} for the two-dimensionalization of incompressible rotating flows. We now take the (projected) circuits in the compressible CBF to be any of the velocity streamlines \textit{screwing} on the cylindrical surfaces and probably closing at infinity or at the periodic boundary with finite circulation in the \textit{stream-screw scenario},\cite{Betchov61} or, to be any of the horizontal vorticity loops binding the vertical ones in the \textit{vortex-knot scenario},\cite{Moffatt78Book} both caricatured clearly in Fig. \ref{fig:ou}: $\mathcal{A}$ changing little geometrically demonstrates that the screws and knots ``fasten'' the gas. Indeed, $\mathcal{A} \to 0$ $\forall$ $c(t)$ results in
\begin{eqnarray}
\text{$\partial_z \bm{u}_h \to \bm{0}$ and $\nabla_h \cdot \bm{u}_h :=\partial_x u_x + \partial_y u_y \to 0$,}  \label{eq:cTPE}\\
\text{but with no explicit constraint on $\partial_z u_z$.} \label{eq:cTPEuz}
\end{eqnarray}
The latter half of Eq. (\ref{eq:cTPE}) implies that the horizontal compressibility $\nabla_h \cdot \bm{u}_h$ may be reduced by (finite) rotation, even for a time-dependent flow (Ref. \onlinecite{Zhu19TaylorProudman}, with analogues for plasma flows and extensions for flows of dimensions $d>3$). $\partial_z u_{z}$ is not supposed to alter so much as to fully compensate the divergence loss of the horizontal flow; and, if the original large compression/expansion lies in the space perpendicular to the rotating plane, enormous reduction of compressibility (thus the heating rate \cite{Schlickeiser2002Book}) may result.

$\mathcal{H}\ne 0$ is thus transformed to the rotation, through Eq. (\ref{eq:Omega}), responsible for `fastening' the above CBF with TPE. In many discussions with a background rotation $\bm{\Omega}_0$ (e.g., \citet{PouquetMininniPTRSA10}), replacing the $\bm{\Omega}$ in the right hand side of Eq. (\ref{eq:hhprime}) whose spatial integral vanishes, whatever the value of $\bm{\Omega}_0$ does not change the helicity,  
which suggests we should avoid the index `$0$' in the CBF $\bm{\Omega}$ (and similarly $\bm{B}$). The \textit{persistence} or \textit{residue} of the effects of such a mechanism may actually be the cornerstone of the helicity fastening principle of turbulence: The flows present disorder, and the (g)CBF(-like) structures may `emerge coherently' (c.f., the remark on \textit{coarse-graining} below for this last phrase). Thus, with the ubiquity of CBF (see below), a CBF ensemble is supposed to be capable of representing the specific statistics of the turbulence in terms of energy and helicity spectra. 
Such a scenario may be viewed as the first step towards an \textit{equivalence principle} of chiral turbulence ensemble, the possibility of which was speculated (\onlinecite{Zhu18PoF2C2Dcw1C3D}, footnote 6) from the \textit{genericity} of the 2C2Dcw1C3D structure related to the fact that the gradient of a vector field is of \textit{real Schur form} up to a local orthogonal transformation acting as an (improper) rotation, an element in the so-called $O(3)$ transformation group (\textit{c.f.}, Ref. \onlinecite{Zhu18PoF2C2Dcw1C3D} and references therein.) Vanishing or any particular value (from \textit{Galilean transformation}) of local helicity density should in general happen in isolated locations, with local 2C2Dcw1C3D structures being ubiquitously helical, thus local CBFs. [Thus, as a side note, we remark that, for the local dynamics, if the local CBF structures could be treated reasonably isolated (a big `if') and if the above transformation via boost to the rotating frame is not for a virtual, merely for simply picturing some more complex scenario effectively, but actual realization, the horizontal compressibility in the local real Schur frame would be statistically smaller than the vertical one. More definitely, choosing for instance the variance as a meaningful statistical measure, we would expect, under the conditions stated, $$\langle |\partial U/\partial X|^2 \rangle + \langle |\partial V/\partial Y)|^2 \rangle)/2 <
\langle |\partial W/\partial Z|^2 \rangle,$$ where $X$ and $Y$ are the horizontal coordinates of the local real Schur frame and $Z$ the vertical one, and, $U$, $V$ and $W$ the corresponding velocity components.]

The idea of CBF ``ensemble'' is related to the `randomness' (varying in space and time) in the properties of the CBFs and $O(3)$ transformation elements, even for a single isotropic turbulent flow. How coherent (g)CBFs emerge and form an equivalent chiral ensemble requires extra physical considerations (see below). The reduction to the CBFs' mean rotations, rather than the mean velocities analogous to Eq. (\ref{eq:Omega}), is due to the consideration of nonrelativistic \textit{Galilean invariance}. For helicity dynamics, the boost to rotating frame is nontrivial, while the Galilean transformation is trivialized by the Newtonian mechanics. This is akin but can be different to the situation where Kraichnan introduced the \textit{random Galilean transformation}, which ``is constant in space and time and has a Gaussian and isotropic distribution over an ensemble of realizations'',\cite{K64RGT} while addressing the large-scale convection/sweeping issue (see also Frisch \cite{Frisch95Book} and references therein for plenty of discussions of \textit{random Galilean invariance}). The randomness of the Galilean transformation, with the velocity determined by an equation analogous to Eq. (\ref{eq:Omega}) as mentioned in the above, mainly lies in that of the formation and property of `coarse-grained' (see below for more definite meaning) CBF itself, thus could also appropriately depend on space and time. 
One however might be concerned by the randomness associated with the scenario that the turbulent fields might even be (multi-)fractal in the inertial range or at infinite Reynolds number (see, e.g., Frisch\cite{Frisch95Book}), thus presumably with asymptotically broken invariance laws (particularly here, the Kelvin theorem and its analogues: c.f., Ref. \onlinecite{EyinkPLA2007} and references therein) that would strictly hold in the ideal flows. Such a consideration does not challenge the relevance of our discussions, because, for instance, in the previous discussions, we can take the material circuits to be initially nice (smooth or sufficiently ``regular" to ensure the applicability of the Kelvin theorem in the Taylor-Proudman effect argument), guaranteeing the relevance of the physics derived from the invariance laws with the necessity of space-time transition intervals to possibly reach an asymptotic deteriorating rough/singular state. Also, the 2C2Dcw1C3D property should hold locally almost everywhere, since possible the singular point set (where the solution is ``weak") must be of dimension less than 3, in the sense that the 3-space volume is zero (otherwise would result in infinite dissipation rate), at each time. In the multi-field case, such as MHD and beyond, the gradients of different fields require their own (improper) rotations to reach the real Schur form, thus we may say that the structures at different locations $\bm{r}$ are `similar' up to $O(3) \times O(3) \times ...$ with each rotation of each field depending on the location, a kind of `gauge similar structure'. Extending the above HD discussions to plasma fluids, with the analogues of the TPE,\cite{Zhu19TaylorProudman} is straightforward, as already indicated in Fig. \ref{fig:ou} for the case of aligned mean magnetic field $\bm{B}$ (for MHD: otherwise, for XMHD and TFPM, the mean field corresponding to the would-be ideally fronzen flux associated to the invariant generalized helicity); see Appendix \ref{app:MHD} for a brief summary of the extension. 

Constructing from the CBFs the solutions of turbulence with some equivalence principle for particular purposes is possible, but the substantiation is beyond the analysis interest of this note. It is not the current purpose to use phenomenological model for some quantitative values, but such an objective is encouraged by the universality and genericity of CBF structures remarked in the above. For example, for isotropic turbulence, a crude rudimentary model ensemble can be constituted by (statistically\cite{ft:statisticallyIdentical}) identical CBFs equally distributed along all directions, each CBF characterized by the energy and helicity spectra, and the turbulence to be an `ideal gas' of CBFs. That is, such a turbulence model can be accurate at the level of energy and helicity. The underlying `equivalence principle' hypothesis here is that `gauge similarity' of local real Schur field structures can be effectively replaced by (statistical) `gauge identity' or much stronger similarity (allowing still other possible freedoms of differences that we are not interested in here) of (g)CBFs, together with the removal of the interactions. 

To be more definite and clear, we may write down the following equivalence principles, roughly resembling those of general relativity (GR),\cite{WeinbergGravitationCosmology} though the analogy is not close: our purpose is for bridging the `fastened' property of the locally ubiquitous helical real Schur flow/field structure, CBF, with that of the globally helical turbulence, not for the geometrization of some force field (in the local transformation, which indeed geometrizes the flow dynamics but is not our current interest,\cite{ft:geometrohydrodynamics} between the laboratory and the local real-Schur coordinates) as the gravitation field in GR. Thus, a \textit{weak} equivalence principle is proposed with the statement that \textit{the statistical properties at the level of energy and helicity (spectra) of a turbulence can be represented by the CBFs with the same energy and helicity as those of the former.} And, a \textit{stronger} equivalence principle would be that \textit{other turbulence statistics that we are interested in, besides the energy and helicity, can also be represented by the CBFs}. We do not expect a strongly universal equivalence principle, but rather the dependence on different purposes. We need only the weak one in this note, and further details of the distributions for, say, an isotropic turbulence, can be equipped with isotropic distribution of the orientation of CBFs and identical amplitudes of the energy and helicity spectra, sufficing for a kinetic description at such a level. 

The phenomenology of the specific simple (minimal) CBF ensemble theory for compressible helical turbulence outlined in the above is dynamically favorable: For homogeneous turbulence, even though the local helicity density $\bm{\omega}\cdot \bm{u}$ varies over positive and negative values, each effective CBF may be understood to be some kind of ``coarse-graining'' (not necessarily simply over local spatial domain, but could be over some phase space of some parameters) of the original field with minute variations of the helicity densities around the global value $\mathcal{H}$, thus it is reasonable to assume an ensemble of such CBFs for our statistical geometrical fluid mechanics. For clarification, we reiterate that ``coarse-graining" here merely means a way to extract or present the effective CBFs from the self-organization transforming the `gauge similarity' to `gauge identity'; thus, it could be simple local spatial averaging or filtering, modeling the possible scenario of particular local CBFs of opposite signs of helicities grouping together to form gCBFs of helicity $\mathcal{H}$, reminiscent of the in (very) large eddy simulations, but could also be other operations in some phase space to parameterize the self-organizing structures. For example, it may be possible to group the Fourier modes in such a way that each sub-set forms a CBF, and all these CBFs are of helicity close to $\mathcal{H}$ and of weak mutual interactions. \textit{A corresponding systematic approach would be to design an appropriate CBF (of helicity $\mathcal{H}$) representation/decomposition technique for the (helical) field.} It is not impossible to go even further to describe nonhomogeneous turbulence with some model of CBF dynamics. Complexification of the ensemble to be more realistic may also be made by enriching the distribution of the properties of the CBFs and their interactions.

\subsection{Plasma statistics and geometries within and beyond compressible XMHD}\label{sec:plasma}

The essential Fourier-mode-interaction and mechanical/geometrical arguments for the fastening notion in the last two sub-sections do not completely exclude the relevance with laminar flows (such as the stability problem) but should be more effective in turbulence where the interactions and CBFs are more active. And, with the fluid mechanical/geometry, it is not surprising that the fastening effect may work \cite{HTG,ClarkDiLeoniMininniBrachetPRA16} reasonably beyond the very restrictive conditions assumed for the statistical calculations for HD.\cite{K55,Z16} Some relevant geometry of plasma fluids having already been remarked, especially in the Appendix \ref{app:MHD}, we proceed to offer further statistical results and extra geometrical insights.

\subsubsection{\label{sec:compressibleXMHD}Compressible XMHD}
Besides Eqs. (\ref{eq:adiabatic},\ref{eq:k1}), the XMHD equations in Alfv\'enic units read
\begin{eqnarray}
\partial_t \bm{u}  = -\nabla \left[\Pi +  u^2/2+(d_e\nabla \times \bm{b})^2/(2\rho^2)\right] +\nonumber\\
+  \bm{u} \times\left(\nabla \times \bm{u}\right)+ \left(\nabla \times\bm{b}\right) \times \bm{\ddot{b}}/\rho, \label{eq:X1}
\end{eqnarray}
\begin{eqnarray}
\partial_t \bm{\ddot{b}} = \nabla \times \left(\bm{u} \times \bm{\ddot{b}} \right) - d_i \nabla \times \left[\left(\nabla \times\bm{b}\right) \times \bm{\ddot{b}}/\rho\right] +\nonumber\\
+ d_e^2 \nabla \times \left[\left(\nabla \times \bm{b}\right) \times \left(\nabla \times \bm{u}\right)/\rho\right], \label{eq:X2}
\end{eqnarray}
with $d_i$ and $d_e$ being the ion and electron skin depths, $\Pi$ the enthalpy, and, for notational convenience,
\begin{equation} \label{eq:X3}
\text{$\bm{\ddot{b}}:=\bm{b}+d_e^2\nabla \times (\nabla \times \bm{b})$, $\nabla \times\bm{\ddot{a}}=\bm{\ddot{b}}$.}
\end{equation}

The Hamiltonian structure, the geometrical formulation with two Lie-carried flux objects and topological aspect connected to the Chern-Simons theory, all of which unify different reduced models (such as the classical single-flow MHD, electron MHD, inertial MHD and Hall MHD) within this framework, have been nicely reviewed in Ref. \onlinecite{MiloshevichLingamMorrisonNJP17} where the incompressible absolute equilibria have been calculated as well. Here we focus on the reduction of compressibility, i.e., the fastening notion.

We can check that Kraichnan's examination of the system and the final computation of absolute equilibrium analysis for HD, briefed in Sec. \ref{sec:statisticsRetrospect}, apply also to XMHD, \textit{mutatis mutandis}: 
We still have Eq. (\ref{eq:zeta}) for the internal energy but should
\begin{equation}\label{eq:xmhdE}
\text{add $\frac{\langle b^2 + d_e^2(\nabla \times \bm{b})^2 \rangle}{2} = \sum_{\bm{k},s}\frac{ \hat{\ddot{b}}_{\bm{k}}^s }{2(1+d_e^2 k^2)}$ to $\mathcal{E}$} 
\end{equation}
in Eq. (\ref{eq:ce}) for the (approximate) XMHD energy (with $\rho_0 = 1$). And, corresponding to Eq. (\ref{eq:ch}), there are two indeally invariant XMHD helicities, $\mathcal{H}_M := \langle \ddot{\bm{a}}\cdot \ddot{\bm{b}}+d_e^2\bm{\omega}\cdot \bm{u} \rangle/2$ and $\mathcal{H}_C := \langle 2\bm{u}\cdot \ddot{\bm{b}}+d_i\bm{\omega}\cdot \bm{u}\rangle/2$, thus
\begin{eqnarray}
\mathcal{H}_M = \frac{1}{2}\sum_{\bm{k},s} (skd_e^2|\hat{u}^s|^2+s|\hat{\ddot{b}}^s|^2/k),\\ \label{eq:XHm}
\mathcal{H}_C = \frac{1}{2} \sum_{\bm{k},s} (\hat{u}^{s*}\hat{\ddot{b}}^s+c.c.+skd_i|\hat{u}^s|^2).\label{eq:XHc}
\end{eqnarray}
From the canonical distribution $\sim \exp\{-(\alpha \mathcal{E} + \beta_M \mathcal{H}_M + \beta_C \mathcal{H}_C ) \}$, we have, besides the same Eq. (\ref{eq:VA}) for density and parallel/compressive modes,
\begin{eqnarray}
U_K^{\pm}(\bm{k}) = \frac{1}{
\alpha \pm k(\beta_M d_e^2 + \beta_C d_i)-\frac{
\beta_C^2}{
\frac{\alpha}{1+k^2 d_e^2} \pm \frac{\beta_M}{k}
}
},\label{eq:xmhdUKs}
\end{eqnarray}
Comparing Eqs. (\ref{eq:xmhdUKs}) and (\ref{eq:VA}), we see, mostly easily by letting $\beta_M=0$ but $\beta_C \ne0$, or the other way round, the increase of the proportion occupied by the vortical energy $U_K^{\perp}(\bm{k})$ in the total energy defined in the same way by relation (\ref{eq:neq}), or, in turn, the relative reduction of the flow compressibility. Small-scale physics enter with the $d_i$ and $d_e$ terms accompanied with $k$ factors, indicating the dependence of scales relative to the ion and electron skin depths, $d_i$ and $d_e$. We focus on the universal statistical energy-partition (between solenoidal and compressibility-relevant modes) perperty and will not delve into the sub-ion details (see, e.g., Ref. \onlinecite{MiloshevichMorrisonTassiPoP2018,ServidioETC08,ZhuMNRAS17,XiaCPL2017} for relevant incompressible-case remarks on the reductions to models such as Hall MHD, among others, and next section on the simpler compressible MHD which is more convenient for detailed discussions).

\subsubsection{Compressible MHD}
We now turn to the MHD equations which, besides Eqs. (\ref{eq:adiabatic},\ref{eq:k1}), may be formally viewed as the limit of $d_e \to 0$ and $d_i \to 0$ from XMHD
\begin{eqnarray}
	\partial_t \bm{b} - \nabla \times (\bm{u}\times \bm{b}) = 0, \label{eq:Mfrozenin}\\
	\partial_t \bm{u} + \bm{u}\cdot \nabla \bm{u} + c^2 \nabla \zeta + \frac{ \nabla \times \bm{b} \times \bm{b} } { \rho_0\exp\{\zeta\} } = 0. \label{eq:momentum}
\end{eqnarray}
Now the ideally invariant energy $\mathcal{E}$, magnetic helicity $\mathcal{H}_M$ 
and the cross helicity $\mathcal{H}_C$ 
are simply the limits of $d_e \to 0$ and $d_i \to 0$ from those of XMHD the latter two of which are chosen particularly for such reductions in Ref. \onlinecite{MiloshevichLingamMorrisonNJP17} from all the linear combinations of them. Accordingly, the MHD absolute equilibrium \textit{modal spectra} are reduced from those of XMHD; for example, besides the same Eq. (\ref{eq:VA}) measuring the flow compressibility, we have
\begin{eqnarray}
U_{K}^{\pm}(\bm{k}) := \langle |\hat{u}_{\pm}|^2 \rangle =\,{\frac {4(\alpha\,k \pm \beta_M)}{(4\,{
\alpha}^{2}-{\beta_C}^{2})k\pm 4\,\alpha\,\beta_M}}, &&\ \label{eq:mhdUK}\\
U_M^{\pm}(\bm{k}) := \langle |\hat{b}_{\pm}|^2 \rangle  =\,{\frac {4(\alpha\,k)}{(4\,{\alpha}^{2}-{\beta_C}^{2})k\pm 4\,\alpha\,\beta_M}}, &&\label{eq:mhdUM}
\end{eqnarray}
Since the incompressible-MHD result of Frisch et al. \cite{FrischJFM75} currently a main theoretical argument for dynamo and precisely recovered in the above, is so familiar in the field, we may need to mention that our result is new in the sense that it is derived from the compressible MHD. When, $\beta_M = \beta_C =0$ for vanishing magnetic and cross helicities, we also recover Lee's \cite{Lee1952} equipartition result. And, the relatively simpler structure makes it appropriate to discuss the helicity-effect result a bit more detailly and extensively (touching some dynamo theory, besides the fastening notion) as follows.

If $\beta_C=0$ (thus $\mathcal{H}_C=0$), but $\beta_M\ne 0$ (thus $\mathcal{H}_M\ne 0$), we have (with obviously $\alpha > 0$ for realizability of the energies)
\begin{equation}
U_M := U_M^{+} + U_M^{-} = \frac {2}{\alpha\pm \frac{\beta_M^2}{\alpha k}} > \frac{2}{\alpha} = U^{\sim} = U_{K}^{\perp},
\end{equation}
indicating that the energy partition is favoring the magnetic, but not the kinetic vortical energy $U_K^{\perp}$ nor the compressibility relevant energy $U^{\sim}$ (as the sum of the parallel and density/potential energies), i.e., against all the kinetic and potential energies, especially at small $k$: The dynamo in such compressible case may be easier with more `preys'. 

When $\mathcal{H}_C \ne 0$ (thus $\beta_C \ne 0$),
\begin{eqnarray}
U_{K}^{\pm} > U_{K}^| = Z = 1/\alpha>0, \
\text{thus $U_K^{\perp} > U^{\sim}$}
\end{eqnarray}
even for $\beta_M = 0$ with then $U_M^{\pm} = U_K^{\pm}$. Eq. (\ref{eq:mhdUM}) shows cross-helicity also favors the partition to the magnetic energy. Since both magnetic and solenoidal kinetic modes are strengthened, the only possibility is the reduction of the compressive and density modes, in the fraction sense or with fixed total energy (for normalization), say. The remark on Ref. \onlinecite{ServidioPaSS2007} in the introductory discussion should not be taken as a clear evidence for this so far, and a more careful check is wanted. Comparing the results with vanishing and non-vanishing $\beta_M$, we can infer that the cross-helicity effect should be more pronounced in the case without magnetic helicity, as the larger fraction of the `dynamo' effect of the latter is screened out.

Both the above two cases agree with the compressibility-reduction/fastening notion. If the kinetic and magnetic energy injections are different when examining, say, the MHD cross-helicity effect, other effects may enter, requiring appropriate normalizations for comparisons. 

As emphasized in Ref. \onlinecite{ZhuMNRAS17} within the incompressible XMHD framework concening the nonlinear dynamical chirality and the linear helical waves (see also Ref. \onlinecite{hydrochirality} for relevant remarks in the context of TFPM), confusion with the dynamics of linear waves (Alfv\'enic, fast and slow) should be avoided, but a prediction of nonlocally and nonlinearly reducing also the turbulent magnetoaeroacoustic `noise', extended beyond the aeroacoustic noise discussed in Ref. \onlinecite{Z16}, may be reasonable. Eqs. (\ref{eq:mhdUK},\ref{eq:mhdUM}) also indicate higher efficacy of cross-helicity at larger-$k$ regime where the dissipation of turbulence at very large $k$ however may enter to reduce or enhance such nonlinear effect. The careful remarks in Ref. \onlinecite{Z16} for turbulence (decay and others) in principle should still apply, though the very preliminary indirect and direct numerical results \cite{ClarkDiLeoniMininniBrachetPRA16,HTG} do have verified the explicitly visible persistence of the above mode-interaction effects. Concerning the energy of the electric field given by Ohm's law $\bm{E} = -\bm{v}\times \bm{b}$, it appears that we can not have clear simple spectral formulae for use to conclude the helicity effect (but see remarks below on the two-fluid model). %

\subsubsection{\label{sec:cTF}Compressible TFPM}
In Eqs. (\ref{eq:adiabatic},\ref{eq:k1}), all the variables are indexed with `$e$(lectron)' and `$i$(on)' respectively for the two fluids, thus it would not be appropriate for the acoustic velocities and densities of them to be assigned with unit normalization value (but the speed of light $c$ here can be!) That is, for $\chi=$ `$e$' or `$i$',
\begin{eqnarray}
\text{$p_{\chi}=c_{\chi}^2\rho_{\chi}$ and $\rho_{\chi}=\rho_{\chi0} e^{\zeta_{\chi}}$
},\label{eq:TFadiabatic}\\
\partial_t \zeta_{\chi} +\zeta_{\chi,\alpha}u_{\chi\alpha}+u_{\chi\alpha,\alpha} = 0. \label{eq:TFcontinuity}
\end{eqnarray}
Similarly are the momentum equations, in which the electric and Lorentz force present: The inviscid version of Eq. (\ref{eq:k2}) includes additionally
\begin{equation}\label{eq:TFmomentum}
\text{$\frac{\rho_{\chi} q_{\chi}}{m_{\chi}}(\bm{E} + \bm{u}_{\chi}\times \bm{b})$,}
\end{equation}
where $\bm{E}$ is the electric field and $m$ the mass, $q$ the electric charge. 
The Maxwell equation system (see below) instead of whatever (generalized) Ohm's law is directly coupled. This is the most comprehensive model for us to study the universality of the fastening effect of helicity in plasmas, which however does not indicate that it is complete (as will be remarked concerning the relativistic effect.) 

\paragraph{\label{prg:cTFstatistics}Statistics}
We can check that all the considerations for the absolute-equilibrium analysis presented in Sec. \ref{sec:statisticsRetrospect} assure it eligible to formally carry out the corresponding calculations. Now the hypothesized absolute equilibria
are given by the canonical distributions $\sim \exp\{-(\alpha \mathcal{E} + \beta_e \mathcal{H}_e + \beta_i \mathcal{H}_i ) \}$ with the constraints of the (approximate) invariant energy 
\begin{equation}\label{eq:tfEnergy}
\mathcal{E} = \langle \bm{E}^2 + \bm{b}^2 + \sum_{\chi}(\rho_{0\chi}\bm{u}_{\chi}^2 + c_{\chi}^2 \zeta_{\chi}^2) \rangle/2
\end{equation}
and the self-helicities, defined by the canonical momenta $\bm{P}_{\chi} := m_{\chi} \bm{u}_{\chi} + q_{\chi}\bm{a}$,
\begin{equation}\label{eq:tfHelicity}
\text{$\mathcal{H}_{\chi} := \langle \nabla \times \bm{P}_{\chi} \cdot \bm{P}_{\chi} \rangle/2 \rangle$ for $\chi = i$ and $e$}, 
\end{equation}
which can be represented, similar to Eqs. (\ref{eq:ce},\ref{eq:ch}), in the Fourier space with the Helmholtz and helical decompositions, resulting in, among others, the equipartitioned
\begin{eqnarray}
Z_{\chi}(\bm{k}):= \langle c_{\chi}^2|\hat{\zeta}_{\chi}|^2 \rangle = \frac{1}{\alpha}= \langle \rho_{0\chi}|\hat{u}_{\chi |}|^2 \rangle =: U^|_{\chi K}(\bm{k}),\ \ \ \label{eq:tfCspectra}\\
\text{$\langle |\hat{E}|^2 \rangle = \frac{1}{\alpha}$ or the finner $\langle |\hat{E}_\pm|^2 \rangle = \langle |\hat{E}_||^2 \rangle = \frac{1}{\alpha}$}.\ \ \ \ \label{eq:electricField}
\end{eqnarray}
Note that the two relations in Eq. (\ref{eq:electricField}) should be used exclusively, not simultaneously, and they do not contradict: It is only that the relation between $\alpha$ and the global invariants Eqs. (\ref{eq:tfEnergy},\ref{eq:tfHelicity}) are quantitatively different for the two relations. 

We have resisted writing in detail all the solenoidal-modal spectra (see Ref. \onlinecite{hydrochirality} for results of the incompressible two-fluid model), because the above equations in the fashion similar to those in other one-fluid models should be sufficient enough to indicate (and is easy, though tedious, to check) the solenoidal modes of magnetic and velocity fields be relatively enhanced by the self-helicities, or the compressive and density/internal-energy modes of the two fluids be reduced, just like the neutral gas flows discussed in Sec. \ref{sec:statisticsRetrospect}. What is particularly new is that the electric field fluctuations are also implied in the similar way to be relatively reduced. Of course, the definite effects in realistic situations (such as the cosmic ray acceleration) should be treated more carefully with now in the dynamics more working actors whose times scales determine the physical relevance. What our somewhat abstract analysis is merely to indicate the possibility and the fundamental mechanism.

\paragraph{\label{prg:cTFgeometry}Geometry}
The geometrical scenario exposed in Sec. \ref{sec:geometry} and Appendix \ref{app:MHD} for helicity reducing flow compressibility associated to the Taylor-Proudman theorem and its analog appears generic and obviously suitable, with appropriate extensions, for more complex fluids, including the XMHD and two-fluid plasmas. There however appears to be no such or other direct geometrical mechanism responsible for the reduction of electric field fluctuations by the self-helicity, which, if indeed, requires us to make connection to the compressibility (reduction).  The explicitly $\bm{E}$-related Maxwell equation with clear geometrical picture is the Gauss law, besides $\nabla \cdot \bm{b} = 0$,
\begin{equation}\label{eq:Edivergence}
\text{$\nabla \cdot \bm{E} = \sum_{\chi} \frac{q_{\chi}}{m_{\chi}} \rho_{\chi}$ or $\hat{E}_| = -\hat{i}\sum_{\chi} \frac{q_{\chi}}{m_{\chi}k_|} \hat{\rho}_{\chi}$},
\end{equation}
stating (with unit electric dielectric coefficient in the appropriate units and scales) that the electric displacement flux out of a surface enclosing a volume is the sum of the linear combination of masses (interpreted from the particle number densities for computing the total charge densities). If the density fluctuations of the two fluids are (approximately) independent, as is the case for the absolute-equilibrium ensemble, it is directly seen that the variance of $\bm{E}_|$ is accordingly also the linear combination of those of the former. In other words, the fastening effect on the density fluctuations, i.e., the geometrical confinement on mass variation, presents directly in $\bm{E}_|$, and the nonlinear two-fluid dynamics tend to equi-partitionize such an effect through the thermalization process. This way, the fastening scenario presented in Sec. \ref{sec:statisticsRetrospect} (now generalized to TFPM) is then understood to be working also on the electric field.

The \textit{neutrality condition} $\nabla \cdot \bm{E} = 0$ follows the infinite-speed-of-light approximation in XMHD and its reductions; thus, the above understanding could reveal in such models, only if those (generalized) Ohm's laws somehow reflect the mechanism. And, we can derive, from Faraday's and Ampere's laws, the equations for $\bm{b}$ and $\bm{E}$ with acoustic/wave operators on the left hand sides and other ``source'' terms (which are not independent on the left-hand-side variables, but just follow Lighthill's\cite{Lighthill1952} `acoustic-analogy' terminology) on the right-hand sides, as well as similar ones for the densities, $\rho_{\chi}$, with now the `Lighthill stress tensor' being more complicated involving extra electromagnetic fields. Lighthill's acoustic analogy has become the major approach to aeroacoustic problems, but the turbulence noise issues are nonlinear and so subtle, both theoretically and numerically, that they are in general not trivially trackable by such (quasi-)linear approach. \cite{ProudmanPRSLA1952,CrowSiAM1970,WangARFM2016,K53} Indeed, preliminary analysis appears to indicate that our helicity-fastening-effect result is not directly obtainable from such an acoustic analogy approach, details of which however is not the focus of this note and will be communicated elsewhere.

\section{Further discussions}\label{sec:AstroImplic}

The background guiding field $\bm{B}_0$ in MHD removes $\mathcal{H}_M$, but not $\mathcal{H}_C$, from the conservation laws.\cite{MatthaeusGoldsteinJGR1982StriblingMatthaeusGhoshJGR1994,ShebalinJPP06} The relevance of magnetic helicity in the presence of $\bm{B}_0$ depends on how approximately $\mathcal{H}_M$ is preserved or the time scale of the term involving $\bm{B}_0$ that breaks the $\mathcal{H}_M$ invariance, which in a sense emphasizes the cross-helicity effect. It is then reasonable to expect applications in much more complex realistic situations, with all the caveats relevant to the assumptions and approximations, and, other subtleties such as the anisotropy issue, kept in mind. Actually, even when a solid-body rotation $\bm{\Omega}_0$ presents with $\sigma \bm{B}_0 = \bm{\Omega}_0$ and a real number $\sigma \ne 0$, a `parallel helicity' $\mathcal{H}_P = \mathcal{H}_C - \sigma \mathcal{H}_M$ is invariant,\citep{ShebalinJPP06} which makes Eqs. (\ref{eq:mhdUK},\ref{eq:mhdUM}
) and others, with $\beta_M=-\sigma \beta_C$ and $\beta_C$ replaced by $\beta_P$, still relevant and similar $\mathcal{H}_P$ fastening effect favorable:
\begin{equation}\label{eq:HPdynamo}
Q_M^{\pm}(k)  := \pm \frac{1}{k}U_M^{\pm}(k) = {\frac {\pm 4\alpha}{(4\,{\alpha}^{2}-{\beta_P}^{2})k\mp 4\sigma \alpha \beta_P}},
\end{equation}
for the modal spectral density of each chiral sector of magnetic helicity, indicates a possible dynamo due to $\mathcal{H}_P$, in a way argued by \citet{FrischJFM75} and \citet{PouquetJFM76} for the magnetic helicity inverse cascade (c.f., Ref. \onlinecite{hydrochirality} for the correspondingly updated discussion with such chirally decomposed finer structures). And, this dynamo will also take energy from the compression and density (or pressure) modes to fasten the flow. 
Similar discussions apply to XMHD and TFPM. Note that the background fields introduce linear terms, thus waves, but no change to the structures of the original quadratic convolutions of the modes, thus the dynamical properties of the latter are still all there; so, the background fields, just like the internal damping effects, add complexities but do not completely sweep the physical results discussed in the last sections out.

The screws work better than nails in joining, say, two plates, and the well winded or knotted ropes bind things better than arbitrary tangles. Fluids seem to be more tameless to be fixed as solids. However, our analyses show that the flow actually may be similarly subject to the notion of ``tightened up'' by the helicity, in the sense of bearing less proportions of compressibility relevant fluctuations such as the compressive, density and internal-energy modes, which can be associated via boost to a rotating frame to the mechanics or geometry in the TPE, with a scenario resembling the situation for solids. The magnetic field, proportional to the gyrofrequency of charged particles, can be regarded as a kind of macroscopic vorticity of the microscopic gyromotions, allowing a unified view of all kinds of helicities for the multidisciplinary \textit{universal law} of `fastening the gases'. Even more systematic quantification of such a fastening notion, particularly the specific \textit{equivalence principle} concerning the CBF ensemble, would be promising for understanding various relevant issues.

However, we should add a caveat as follows. Since direct analytical solutions of turbulent flows are not available, one relies either on phenomenological or statistical arguments. Such arguments may be considered as heuristics, in particular, because we know that absolute equilibrium states do not realize themselves in realistic systems with the forward cascade so that conclusions directly from one particular type of them must be taken with a grain of salt. Instead, we have made predictions from the comparisons of helical and nonhelical absolute equilibria, not merely from one of them, and we have also found connections with the very fluid geometry/mechanics, in which sense the results may deserve serious treatment. Nevertheless, plasma dynamics are very complicated with mixed ingredients not fully included in our statistical and geometrical calculations, and it is not impossible that other effects may completely change even the qualitative feature presented in the absolute equilibria, making the comparison of the latter be of little relevance or making the derived effect from the comparison very difficult to be extracted (in the anisotropic case with strong background field and/or with weak nonlinearity, say). Thus, whatever the hypothesis and prediction in this note should be further testified carefully with details for further clarifications, as preliminarily done in Ref. \onlinecite{HTG}, with more rigorous numerical checks being initiated.

Ideal conservations of circulations and helicities also present in the relativistic flows,\cite{MarkakisEAprd2017} however, the reduction to rotation might not be complete for lack of the Galilean invariance (Sec. \ref{sec:geometry}), indicating the requirement of more general considerations for a fundamental relativistic statistical geometrical fluid mechanics. This is important also in practical usage, because, though many flow noise or stability problems in civil engineering, rapid public transportation and car transportation, and, aeronautics, say, are well nonrelativistic, others, in particular of plasmas flows, in controlled fusion or astrophysics or space physics (such as the cosmic ray and solar physics problems mentioned in Sec. \ref{sec:introduction}) require considering the (extremely) relativistic effects. The other promising branch for extension and/or deeper exploration of the relevant ideas is the quantum flow where other aspects of helicity may be revealed by the (macroscopic) quantum effects, as also partly indicated in the introductory discussions and anticipated from the relevant analyses of neutral superfluids and superfluid plasmas,\cite{SoninRMP1987,HolmKupershmidtPRA1987} being obviously of high-tec industry and astrophysical (neutron stars, say) applications as well. 

\begin{acknowledgements}
Supported by NSFC (No. 11672102) and the Ti\'an-Yu\'an-Xu\'e-P\`ai (No. 27182818) foundation. The referees helped improve the presentation and Wennan Zou imparted Ref. \onlinecite{LiZhangHeAPS2014}, which together with many other questions and suggestions from colleagues are warmly acknowledged. 
\end{acknowledgements}

\section*{Data availability}
The data that support the findings of this study are available from the corresponding author upon reasonable request.

\appendix

\section{Magnetic helicity, relative vector potential and mean magnetic field}\label{app:MHD}
First, we extend the analogue of TPE for MHD with a strong background uniform magnetic field $\bm{B}$ to the compressible case. In the incompressible case, \citet{Chandrasekhar61Book} presented it in the `popular' way (completely abandoning the geometrical argument of \citet{TaylorPRSA21}), without invoking the geometry of the Alfv\'en theorem as the analogue of the Helmholtz-Kelvin theorems, with the result that $(\bm{B}\cdot \nabla) \bm{u} = 0$: for $\bm{B}$ lying in the $z$ axis, $\partial_z \bm{u} = \bm{0}$. Now, for the compressible but the barotropic case, with $\bm{a}$ (and $\bm{a}'$) in the Alfv\'en theorem 
\begin{equation}
\text{$\oint_{c(t)} \bm{a}\cdot d\bm{r} = \int_{\partial^{-1}c(t)} \bm{b}\cdot d\bm{s}=const.$}
\end{equation}
replaced with $\bm{u}$ (and $\bm{u}'$) in Eq. (\ref{eq:circulation}), and $B$ with $\Omega$ there, the same geometrical argument of \citet{TaylorPRSA21}, with $\bm{A}:=(\nabla \times)^{-1} (2\bm{B}) = (-yB,xB,0)+\nabla \varphi$ whose gradient of potential $\varphi$ does not contribute to the loop integral, leads to the same compressible TPE formula as in Eq. (\ref{eq:cTPE}). Alternatively, in the analysis of \citet{Chandrasekhar61Book}, an extra term $\bm{B}(\nabla \cdot \bm{u})$ transforms his Eq. (83) to our Eq. (\ref{eq:cTPE}). Actually, the treatment in terms of differential forms,\cite{Zhu19TaylorProudman} extending TPE to compressible and higher-dimensional cases with the geometric notion, carries over to the magnetic field corresponding to a 2-form $\verb"B" = \verb"d"\verb"A"$ (thus the projected area in Taylor's argument naturally emerges). The emphasis here however is the geometrization of the fastening notion with such geometrical `screws' and `ropes' in the TPE mechanism:---

Now for the magnetic case, with the two-dimensionality of the horizontal potential $\partial_z \bm{a}_h = \bm{0}$ and the CBFs assumed to have the properties similar to those in Sec. \ref{sec:CBFrot}, we can introduce $\bm{B}$, from nonvanishing magnetic helicity $\mathcal{H}_M$, as $\bm{\Omega}$ in Eq. (\ref{eq:Omega}). Thus, the above analogue of compressible TPE indeed geometrizes the fastening effect of $\mathcal{H}_M$ for such a CBF. $\bm{\Omega}$ shall not be aligned with $\bm{B}$ when we have the original cross-helicity non-vanishing but want to make that of the relative fields ($\bm{u}'$ and $\bm{b}'$) vanish. Alfv\'en's theorem again provides the geometrization of the fastening effect with the compressible TPE and its analogue. We note that in our CBF turbulence ensemble, the CBFs should correspond to the dynamically important transient state of the relaxation, the end of which, however, points to (anti-)alignment between $\bm{u}$ and $\bm{b}$ (Alfv\'enicity). The two states are different, though the (local) relaxation can be fast \cite{MatthaeusEAprl2008} and a mixture of them may appear in various complex realistic situations (e.g., Ref. \onlinecite{TuMarschSSR1995} for a review in space science). 

Just to reiterate, we have transformed the helicity-effects (with or without $\bm{B}_0$ and/or $\bm{\Omega}_0$: Sec. \ref{sec:AstroImplic}) to pure $\bm{B}$-and/or-$\bm{\Omega}$-effect one. The generalizations of the above discussions to CBFs of XMHD, or even two-fluid model, are straightforward, with $\bm{u}$ and $\bm{b}$ replaced by the two generalized momentums, whose curls (generalized vorticities) are ideally frozen-in to two flows.

\nocite{*}

\begin{thebibliography}{10}
\bibitem{ft:geometrohydrodynamics}
This would correspond, in the context of the loose analogy with GR, to studying the transformation to eliminate the local connection to obtain the locally inertial frame, which to our best knowledge (c.f., Ref. \onlinecite{WeinbergGravitationCosmology}) has neither been well performed in GR.
\end{thebibliography}

\end{document}